\documentclass[a4paper,12pt]{article}
\usepackage[esperanto,portuguese,english]{babel}
\usepackage{epsfig}
\usepackage{icomma}
\usepackage{float}

\newcommand{\ppparallel}[1]{} \newcommand{\ppmaldekstra}[1]{} \newcommand{\ppdekstra}[1]{}
\usepackage{parallel} \renewcommand{\ppparallel}[1]{#1} 

\newcommand{\comentario}[1]{}
\newcommand{\Sch}{Schwarz\-schild}
\newcommand{\dd}{\mathrm{d}}

\ppparallel{
\newlength{\pplw}\setlength{\pplw}{0.47\textwidth}
\newlength{\pprw}\setlength{\pprw}{0.49\textwidth}

\newcommand{\ppp}{\ParallelPar}
\newcommand{\ppn}{\noindent}              
\newcommand{\ppl}[1]{\ParallelLText{\selectlanguage{esperanto}#1}}
\newcommand{\ppr}[1]{\ParallelRText{\selectlanguage{english}#1}\ppp}
\newcommand{\ppln}[1]
{\ParallelLText{\ppn \selectlanguage{esperanto}#1}} 
\newcommand{\pprn}[1]
{\ParallelRText{\ppn \selectlanguage{english}#1}\ppp} 

\newcommand{\ppsection}[3][0ex]{\vspace{2em} 
\ppl{\section{#2} \vspace{#1}} \ppa \nopagebreak
\ppR{\section{#3}} \ppp \nopagebreak}

\newcommand{\ppsubsection}[3][0ex]{\vspace{2em}
\ppl{\subsection{#2}\vspace{#1}} \ppas \nopagebreak
\ppR{\subsection{#3}} \ppp \nopagebreak}

\newcommand{\bea}{\vspace{-1ex}\begin{eqnarray}}
\newcommand{\eea}{\end{eqnarray}}
}

\ppmaldekstra{
\newcommand{\ppl}[1]{\selectlanguage{esperanto}#1}
\newcommand{\ppln}[1]{\noindent \selectlanguage{esperanto}#1}
\newcommand{\ppr}[1]{\selectlanguage{english}}
\newcommand{\pprn}[1]{\noindent \selectlanguage{english}}
\newcommand{\ppsection}[3][0ex]{\section{#2}}
\newcommand{\ppsubsection}[3][0ex]{\subsection{#2}}
\newcommand{\bea}{\begin{eqnarray}}
\newcommand{\eea}{\end{eqnarray}}
}

\ppdekstra{
\newcommand{\ppl}[1]{\selectlanguage{esperanto}}
\newcommand{\ppln}[1]{\noindent \selectlanguage{esperanto}}
\newcommand{\ppr}[1]{\selectlanguage{english}#1}
\newcommand{\pprn}[1]{\noindent \selectlanguage{english}#1}
\newcommand{\ppsection}[3][0ex]{\section{#3}}
\newcommand{\ppsubsection}[3][0ex]{\subsection{#3}}
\newcommand{\bea}{\begin{eqnarray}}
\newcommand{\eea}{\end{eqnarray}}
}

\title{{ Samtempaj geodezioj \^ce \^Svarc\^sild \ppparallel{\\ Geodesics of  simultaneity in \Sch}}}
\author{F.M. Paiva \\ 
{\small Departamento de F\'isica, Unidade Humait\'a II, Col\'egio Pedro II} \\
{\small Rua Humait\'a 80, 22261-040  Rio de Janeiro-RJ, Brasil; fmpaiva@cbpf.br} 
\vspace{.7ex} \\
A.F.F. Teixeira \\
{\small Centro Brasileiro de Pesquisas F\'isicas} \\
{\small 22290-180 Rio de Janeiro-RJ, Brasil; teixeira@cbpf.br}}

\begin{document}
\selectlanguage{esperanto}
\maketitle
\thispagestyle{empty}

\begin{abstract}\selectlanguage{esperanto}
Samtempa geodezio estas geodezio de spaca tipo kies paroj de najbaraj eventoj estas samtempaj ($g_{0\mu}\dd x^\mu=0$). Tiuj geodezioj estas studataj en ekstera regiono de metriko de \^Svarc\^sild.  

\ppparallel{\selectlanguage{english}
Geodesic of simultaneity is a spacelike geodesic in which every pair of neighbour events are simultaneous ($g_{0\mu}\dd x^\mu=0$). These geodesics are studied in the exterior region of \Sch's metric.} 

\end{abstract}

\ppparallel{
\begin{Parallel}[v]{\pplw}{\pprw}
}

\ppparallel{\section*{\vspace{-2em}}\vspace{-2ex}}   

\ppsection[0.6ex]{Enkonduko}{Introduction}

\ppln{Ni konsideras limhavan regionon de spacotempo, kaj supozas ke tiu regiono enhavas kelkan graviton. Elektante du eventojn en tiu regiono, ni nomas geodezio, linion kiu unuigas ilin, tiel ke la sumigo de infinitezimaj intervaloj, $\dd s$\,, en tiu linio estu minimuma per infinitezimaj varioj de tiu linio, estante fiksataj \^giaj ekstremoj. La spacotempo de relativeca teorio permesas tri tipojn de geodezioj: spacan, nulan, kaj tempan. En geodezio de elektata tipo, \^ciu infinitezima intervalo havas tiun saman tipon.}
\pprn{We consider a finite region of spacetime, and assume that the region contains some gravitation. Selecting two events in the region, we call geodesic, a line which connects them, such that the sum of all infinitesimal intervals, $\dd s$\,, in the line be minimum under infinitesimal variations of the line, maintaining fixed the extremities. The spacetime of theory of relativity allows three types of geodesics: spacelike, null, and timelike. In a geodesic of a selected type, every infinitesimal interval has that type.} 

\ppl{Kompreneble, fizikistoj zorgas precipe pri tempa geodezio (pris\-kribanta mo\-va\-don de maso libere fluganta en gravito) kaj pri nula geodezio (pri\-skribanta movadon de lumo). Sed matematiko konsideras ke la tri tipoj estas same gravaj. Tiu fakto stimulis nin atenti spacajn geodeziojn.}
\ppr{Naturally, physicists occupy mainly with timelike geodesic (describing motion of mass freely flying in the gravity) and with null geodesic (describing motion of light). Mathematics, nevertheless, considers the three types equally important. That fact stimulated us to occupy with spacelike geodesics.}  

\ppl{Se koordinatoj estas lokataj en la spacotempa regiono, speciala klaso de spacaj geodezioj aperas: la samtempaj geode\-zioj \cite{reltemp2}. En ili, \^ciu infinitezima intervalo havas $g_{0\mu}\dd x^\mu=0$\,, estante $\dd x^\mu$ la apartigoj de koordinatoj \cite[pa\^go 350]{Anderson}.}
\ppr{If coordinates are set in the spacetime region, a special class of  spacelike geodesics appears: the geodesics of simultaneity \cite{reltemp2}. In them, every infinitesimal interval has $g_{0\mu}\dd x^\mu=0$\,, with $\dd x^\mu$ being the separation of coordinates \cite[page 350]{Anderson}.} 

\ppl{Geodezio persistas geodezio per \^san\^go de koordinatoj, kaj anka\u u \^gia tipo persistas la sama. Tamen, samtempa geodezio en elektata sistemo povas ne persisti samtempa, kvankam persistas spaca. En \^ci tiu artikolo ni studas samtempajn geodeziojn en la pli ordinara formo de metriko de \^Svarc\^sild \cite{Schwarzschild}.}
\ppr{A geodesic remains geodesic under a change of coordinates, and also its type remains the same. However, a geodesic of simultaneity in a selected system may not remain of simultaneity, although it remains spacelike. In this article we study geodesics of simultaneity in the most usual form of \Sch's metric~\cite{Schwarzschild}.} 

\ppsection[0.6ex]{\^Svarc\^sild}{Schwarzschild}

\ppln{Ni konsideras linielementon}
\pprn{We consider the line element}   
\bea                                                       \label{Sch1}
\epsilon(\dd s)^2=(1-\rho/r)(c\dd T)^2-\frac{(\dd r)^2}{1-\rho/r}-r^2(\dd\theta)^2-r^2\sin^2\theta(\dd\varphi)^2\,,  
\eea
\ppln{estante $\epsilon=\pm1$, kaj estante $\rho:=2Gm/c^2$ la radiuso de \^Svarc\^sild. En \^ci tiu artikolo ni prizorgas nur la eksteran regionon $r>\rho$ kaj ni konsideras nur la ebenon $\theta=\pi/2$\,; do ni uzas}
\pprn{where $\epsilon=\pm1$, and where $\rho:=2Gm/c^2$ is the radius of \Sch. In this article we consider only the exterior region $r>\rho$\,, and we consider only the plane $\theta=\pi/2$\,; we then use}   
\bea                                                       \label{Sch2}
\epsilon(\dd s)^2=(1-\rho/r)(c\,\dd T)^2-\frac{(\dd r)^2}{1-\rho/r}-r^2(\dd\varphi)^2\,.   
\eea 

\ppl{Se intervalo estas de spaca tipo oni uzas $\epsilon=-1$, kaj renomas $\dd s\rightarrow\dd\lambda$. Kaj en linielemento (\ref{Sch2}), kondi\^co de samtempeco $g_{0\mu}\dd x^\mu=0$ implicas $\dd T=0$. Tiuokaze (\ref{Sch2}) plisimpli\^gas al}
\ppr{For a spacelike interval one uses $\epsilon=-1$, and redenominates  $\dd s\rightarrow\dd\lambda$. And in the line element (\ref{Sch2}) the condition of simultaneity $g_{0\mu}\dd x^\mu=0$ implies $\dd T=0$. In that case (\ref{Sch2}) simplifies to}    
\bea                                                       \label{prov}
(\dd\lambda)^2=\frac{(\dd r)^2}{1-\rho/r}+r^2(\dd\varphi)^2\,, 
\eea
\ppln{egala al metriko de spacia sekcio en ebeno $\theta=\pi/2$\,. \^Car (\ref{prov}) ne pendas de $\varphi$, tial en \^ciu geodezio okazas $u_\varphi=$ konst $=:-r_0$. Konsekvence $u^\varphi:=\dd\varphi/\dd\lambda=g^{\varphi\varphi}u_\varphi=r_0/r^2$\,, do}
\pprn{equal to the metric of the spacial section in the plane $\theta=\pi/2$\,. Since (\ref{prov}) does not depend on $\varphi$, in each geodesic it occurs $u_\varphi=$ konst $=:-r_0$. Consequently $u^\varphi:=\dd\varphi/\dd\lambda=g^{\varphi\varphi}u_\varphi=r_0/r^2$\,, so}  
\bea                                                       \label{dphi}
\dd\varphi=(r_0/r^2)\,\dd\lambda\,, 
\eea 
\ppln{kaj (\ref{prov}) reskribi\^gas kiel}
\pprn{and (\ref{prov}) rewrites as} 
\bea                                                        \label{dr2}
(\dd r)^2=(1-\rho/r)(1-{r_0}^2/r^2)(\dd\lambda)^2\,.  
\eea 
\ppln{Tirante $\dd\lambda$ el (\ref{dphi}) kaj lokante \^gin en (\ref{dr2}) oni generas diferencialan ekvacion por samtempaj geodezioj:}
\pprn{Extracting $\dd\lambda$ from (\ref{dphi}) and inserting into (\ref{dr2}) one generates a differential equation for the geodesics of simultaneity:} 
\bea                                                       \label{dif1}
(\dd r)^2=(r/{r_0}^2)(r-\rho)(r^2-{r_0}^2)(\dd\varphi)^2\,.  
\eea
\ppln{Sen perdi \^generalecon, ni supozas ke $r_0$ estas pozitiva.}
\pprn{Without loss of generality, we assume $r_0$ positive.}

\ppl{Ni rememoras la jenan geometrian fakton: se $r_0$ estas distanco de E\u uklida rekto al origino, tial en radiusaj infinitoj de rekto okazas $\dd\varphi/\dd r=\pm r_0/r^2$. Do, farante $r\rightarrow\infty$ en (\ref{dif1}), montri\^gas ke samtempaj geodezioj havas {\it asimptotojn} kies kolizia parametro estas $r_0$.}
\ppr{We remind the following geometrical fact: if $r_0$ is the distance from an Euclidean straight line to the origin, then at radial infinities of the line it occurs $\dd\varphi/\dd r=\pm r_0/r^2$. So, taking  $r\rightarrow\infty$ in (\ref{dif1}), shows that the geodesics of simultaneity have {\it asymptotes\,} whose impact parameter is $r_0$.}

\ppl{Pozitiveco de $(\dd r)^2$ en (\ref{dif1}) okazigas du klasojn de solvo kun $r>\rho$\,:}
\ppr{Positiveness of $(\dd r)^2$ in (\ref{dif1}) produces two classes of solution with $r>\rho$\,:}  
\bea  \nonumber                                                  
1\,: r>r_0>\rho\,; \nonumber\\ 
2\,: r>\rho>r_0\,.\nonumber \\ 
\nonumber 
\eea 
\vspace{-1cm} 
\ppln{Tiuj klasoj estos malkune ekzamenataj.}
\pprn{These classes will be examined separately.}

\ppsubsection[0.6ex]{Klaso 1}{Class 1}

\ppln{En klaso $r>r_0>\rho$\,, la radiusa valoro $r_0$ indikas perihelion, kaj (\ref{dif1}) reskribi\^gas kiel}
\pprn{In class $r>r_0>\rho$ the radial value $r_0$ indicates perihelium, and (\ref{dif1}) rewrites as}  
\bea                                                       \label{kiki}
\dd\varphi=\pm\frac{r_0\dd r}{\sqrt{r(r-\rho)(r^2-{r_0}^2)}}\,. 
\eea

\begin{figure}[h]                                              
\centerline{\epsfig{file=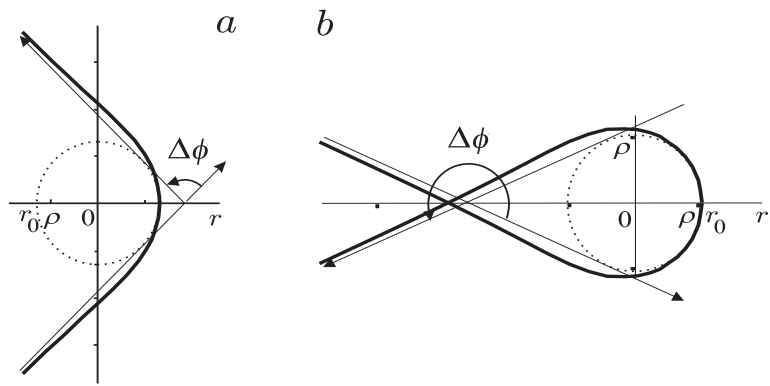,width=10cm,height=5cm}} 
\caption{
Du samtempaj geodezioj (dikaj linioj) de klaso 1\,: $r>r_0>\rho$. Punktitaj cirkloj havas radiuson $r_0$, kaj la maldikaj linioj estas orientitaj asimptotoj. Ili tan\^gas la punktitan cirklon, kaj $\Delta\phi$ estas la angula vario de unu al la alia. En $a$ ni uzis  $r_0=4\rho/3$, farinte $\Delta\phi\approx\pi/2$. En $b$ ni uzis $r_0=1,04\rho$, farinte $\Delta\phi\approx5\pi/4$.
\newline Figure~\ref{FiguraUm}:
Two geodesics of simultaneity (thick lines) of class 1\,: $r>r_0>\rho$. The dotted circles have radius $r_0$, and the thin lines are oriented asymptotes. They tangenciate the dotted circle, and $\Delta\phi$ is the angular variation from one to the other. In $a$ we used $r_0=4\rho/3$,  making $\Delta\phi\approx\pi/2$. In $b$ we used $r_0=1,04\rho$, making $\Delta\phi\approx5\pi/4$.}
\label{FiguraUm} 
\end{figure}

\ppln{La integro de (\ref{kiki}) de $r_0$ \^gis $r$ estas elipsa funkcio; vidu figuron~\ref{FiguraUm}. La geodezio \^cirka\u uas la altirantan centron, kaj ju pli granda estas la totala angula vario $\Delta\phi$ de \^gia tangento, des malpli granda estas la perihelio $r_0$\,.}
\pprn{The integral of (\ref{kiki}) from $r_0$ to $r$ is an elliptic function; see figure~\ref{FiguraUm}. The geodesic surrounds the attractive center, and the bigger is the total angular variation $\Delta\phi$ of its geometric tangent, the smaller is the perihelium $r_0$\,.}

\ppl{Tiu vario estas la elipsa funkcio}
\ppr{That variation is the elliptic function}  
\bea                                                      \label{Delta}
\Delta\phi(r_0)=2\int_{r=r_0}^{r=\infty}|\dd\varphi|-\pi\,, 
\eea 
\ppln{kies grafo estas en regiono $r_0>\rho$ en figuro~\ref{FiguraUmDois}. Tie ni vidas ke $\Delta \phi(\rho) \rightarrow\infty$\,. Do oni montras ke en ajn du punktoj (kun $r>\rho$) trapasas nefinia nombro de samtempaj geodezioj.}
\pprn{whose graph is in the region $r_0>\rho$ in figure~\ref{FiguraUmDois}. There we see that $\Delta \phi(\rho)\rightarrow\infty$\,. So one shows that through any two points (with $r>\rho$) an infinite number of geodesics of simultaneity passes.}

\begin{figure}[h]                                              
\centerline{\epsfig{file=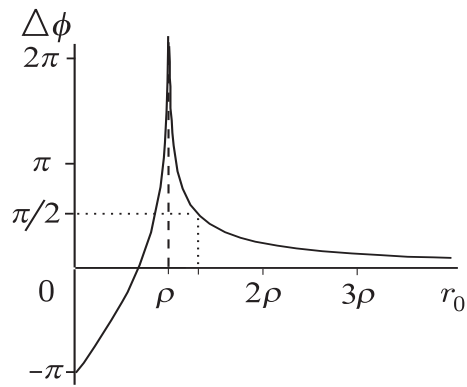,width=6cm,height=5cm}} 
\caption{
Totala vario $\Delta\phi(r_0)$ de oriento de samtempa geodezio de klaso 1\,: $r>r_0>\rho$ kaj de klaso 2: $r>\rho>r_0$. Speciale, la cirklo kun radiuso $r_0=\rho$ estas samtempa geodezio, kun $\Delta\phi(\rho)=\infty$. Ne konfuzu \^ci tiun geodezion kun la nula geodezio anka\u u cirkla, sed kun radiuso  $3\rho/2$\,. Malpozitivaj valoroj de $\Delta\phi(r_0)$ por $r_0<0,67\rho$ indikas ke samtempa geodezio kun tre malgranda kolizia parametro $r_0$ estas forte forpu\^sata de centra maso.
\newline Figure~\ref{FiguraUmDois}:
Total variation $\Delta\phi(r_0)$ of orientation of a geodesic of simultaneity of class 1\,: $r>r_0>\rho$ and of class 2: $r>\rho>r_0$. In particular, the circle with radius $r_0=\rho$ is a geodesic of simultaneity, with $\Delta\phi(\rho)=\infty$. Do not confound this geodesic with the null geodesic also circular, but with radius  $3\rho/2$\,. Negative values of $\Delta\phi(r_0)$ for $r_0<0,67\rho$ indicate that a geodesic of simultaneity with very small impact parameter $r_0$ is strongly repelled from the central mass.} 
\label{FiguraUmDois} 
\end{figure}

\ppsubsection[0.6ex]{Klaso 2}{Class 2} 

\ppln{En klaso $r\!>\!\rho\!>\!r_0$\,, la pozicio $\rho$ estas perihelio. \^Car la kolizia parametro $r_0$ de la asimptotoj estas malpli granda ol  $\rho$\,, tial la {\it asimptotoj} pasas tra la sfero $r=\rho$ de \^Svarc\^sild. Por koni la formon de la geodezioj ni integras (\ref{kiki}) de $\rho$ \^gis $r$\,. Vidu figuron~\ref{FiguraTres}.}
\pprn{In the class $r>\rho>r_0$ the position $\rho$ is perihelium. Since the impact parameter $r_0$ of the asymptotes is smaller than $\rho$\,, the {\it asymptotes} traverse the sphere $r=\rho$ of \Sch. To know the form of the geodesics we integrate (\ref{kiki}) from $\rho$ to $r$\,. See figure~\ref{FiguraTres}.}  

\begin{figure}[h]                                             
\centerline{\epsfig{file=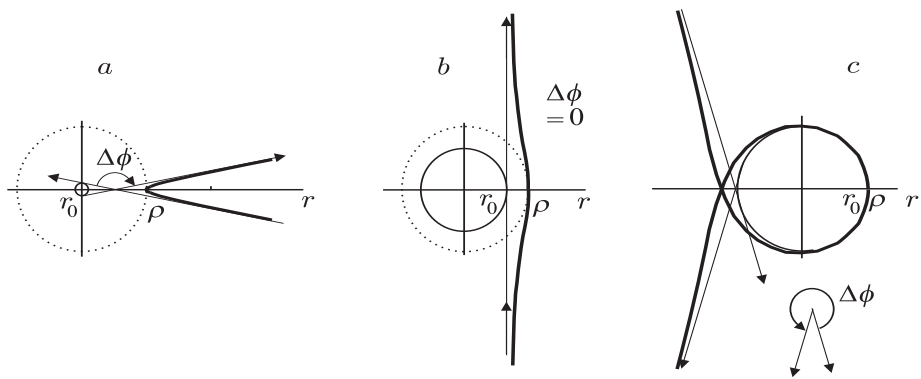,width=12cm,height=49mm}} 
\caption{ 
Tri samtempaj geodezioj (dikaj linioj) de klaso 2\,: $r>\rho>r_0$. La punktitaj cirkloj havas radiuson $\rho$\,, kaj la ne punktitaj havas radiuson $r_0$. La maldikaj linioj estas orientitaj asimptotoj. Ili tan\^gas la cirklon kun radiuso $r_0$\,; la angula vario de unu al la alia estas $\Delta\phi$\,. En $a$ ni uzis $r_0=0,1\rho$, farinte $\Delta\phi<0$; en $b$ ni uzis $r_0\approx0,67\rho$, farinte $\Delta\phi=0$; kaj en $c$ ni uzis $r_0=0,99\rho$, farinte $\Delta\phi>0$\,.
\newline Figure~\ref{FiguraTres}: 
Three geodesics of simultaneity (thick lines) of class 2\,: $r>\rho>r_0$. The dotted circles have radius $\rho$\,, and the non-dotted have radius $r_0$. The thin lines are oriented asymptotes. They tangenciate the circle with radius $r_0$, and $\Delta\phi$ is the angular variation from one to the other. In $a$ we used $r_0=0,1\rho$, making $\Delta\phi<0$; in $b$ we used $r_0\approx0,67\rho$, making $\Delta\phi=0$; and in $c$ we used $r_0=0,99\rho$, making  $\Delta\phi>0$\,.}  
\label{FiguraTres} 
\end{figure}

\ppl{La totala vario de oriento de geodezio estas la integro}
\ppr{The total variation of orientation of the geodesic is the integral}  
\bea                                                     \label{Delta2}
\Delta\phi(r_0)=2\int_{r=\rho}^{r=\infty}|\dd\varphi|-\pi\,, 
\eea 
\ppln{kies grafo estas en regiono $r_0<\rho$ en figuro~\ref{FiguraUmDois}.}
\pprn{whose graph is in the region $r_0<\rho$ in figure~\ref{FiguraUmDois}.}

\ppsection[0.6ex]{Komentoj}{Comments}

\ppln{Bonkonate, la samtempaj geodezioj de spacotempo de Minkovski estas E\u uklidaj rektoj. Tio okazas anka\u u en spacotempo (\ref{Sch1}) de \^Svarc\^sild, sed nur en regionoj kun tre malforta gravito $(r>>\rho)$. Tamen, se la kolizia parametro $r_0$ valoras proksimume la radiuso $\rho$ de \^Svarc\^sild, do la samtempa geodezio tre malsimilas E\u uklidan rekton, precipe en regionoj \^cirka\u u $\rho$. Tio estas klare vidata en figuroj \ref{FiguraUm} kaj \ref{FiguraTres}.}
\pprn{As is well known, the geodesics of simultaneity of Minkowski spacetime are Euclidean straight lines. The same occurs in Schwarzschild spacetime (\ref{Sch1}), but only in regions with very weak  gravitation, $r>>\rho$. However, if the impact parameter $r_0$ is nearly the radius $\rho$ of \Sch, the geodesic of simultaneity differs greatly from an  Euclidean straight line, especially in regions near $\rho$. That is clearly seen in figures \ref{FiguraUm} and \ref{FiguraTres}.}  

\ppl{Pensu pri E\u uklida trispaco, kun linielemento   
$(\dd\lambda)^2 = (\dd r)^2 + r^2(\dd\varphi)^2 + (\dd z)^2$\,, 
kaj pri la paraboloido $z^2=4 \rho(r- \rho)$\,, rotacianta \^cirka\u u akso $z$\,. Uzante tiujn du ekvaciojn, la linielemento de paraboloido estas (\ref{prov}), la sama linielemento kies geodezioj estas la samtempaj geodezioj de \^Svarc\^sild kun $\theta=\pi/2$. Geodezio de klaso 1 rilatas al geodezio de paraboloido kiu ne trairas ekvatoron $r=\rho$ de paraboloido, kaj geodezio de klaso 2 rilatas al tiu kiu trairas. Vidu figuron~\ref{FiguraQuatro2}.} 
\ppr{Consider Euclidean three-space, with line element  $(\dd\lambda)^2=(\dd r)^2+r^2(\dd\varphi)^2+(\dd z)^2$\,,  and the paraboloid $z^2=4\rho(r-\rho)$\,, of revolution around the $z-$axis. Using these two equations, the line element of the paraboloid is   (\ref{prov}), the same line element whose geodesics are those of simultaneity of Schwarzschild with $\theta=\pi/2$. A geodesic of class 1 corresponds to a geodesic of the paraboloid not traversing the equator of the paraboloid, and a geodesic of class 2 corresponds to one that traverses. See figure~\ref{FiguraQuatro2}.}

\begin{figure}[h]                                              
\centerline{\epsfig{file=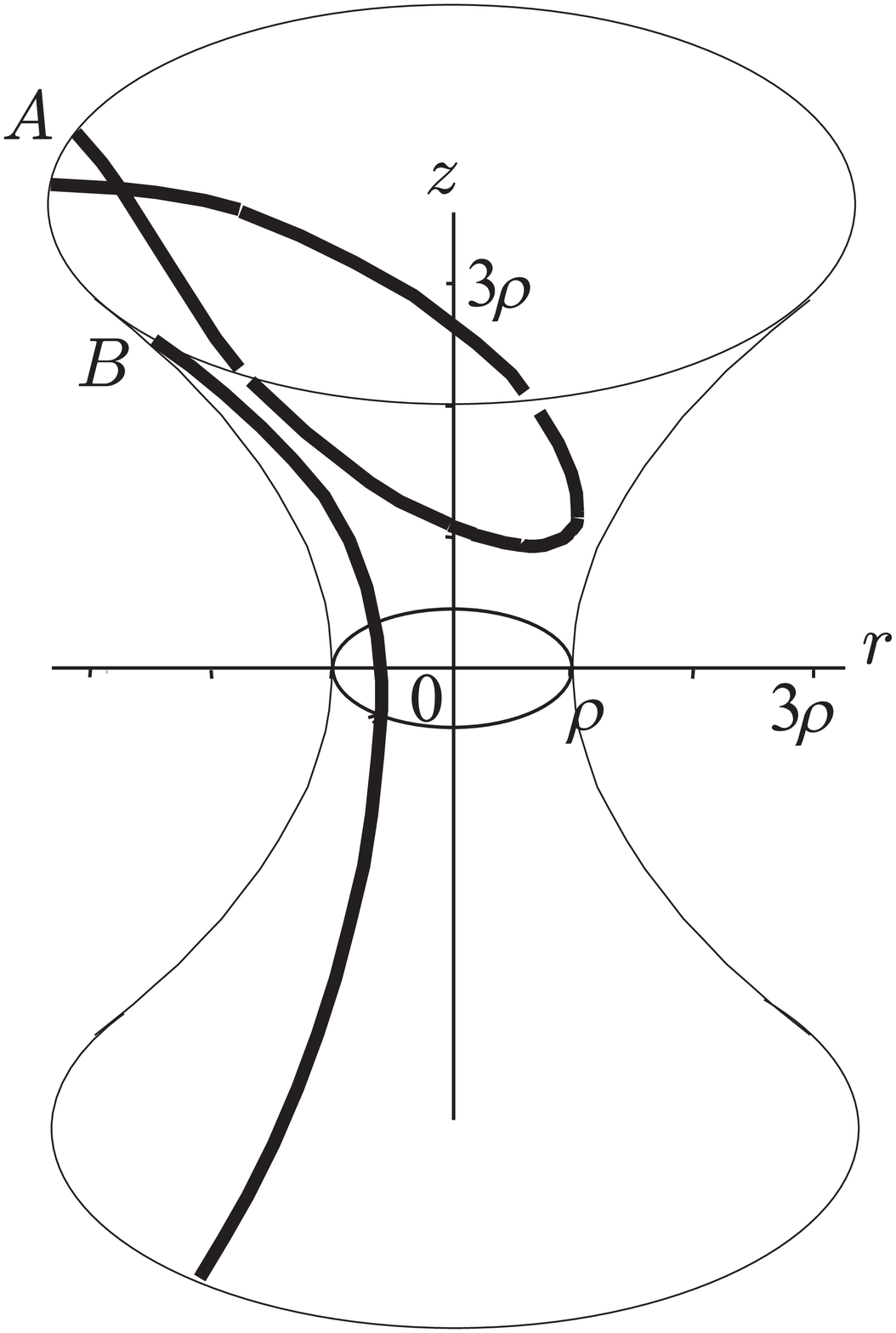,width=5cm}} 
\caption{
Skizo de paraboloido $z^2=4\rho(r-\rho)$\,, kaj du el \^giaj geodezioj. {\it A} estas geodezio ne trairanta ekvatoron $r=\rho$ de paraboloido, kaj rilatas al samtempa geodezio de \^Svarc\^sild en figuro~\ref{FiguraUm}b. {\it B} estas geodezio kiu trairas, kaj rilatas al samtempa geodezio en figuro~\ref{FiguraTres}a.  
\newline 
Figure~\ref{FiguraQuatro2}:
Sketch of paraboloid $z^2=4\rho(r-\rho)$\,, and two of its geodesics. {\it A} is a geodesic not traversing the equator $r=\rho$ of the paraboloid, and relates to a geodesic of simultaneity of Schwarzschild as in figure~\ref{FiguraUm}b. {\it B} is a geodesic that traverses, and relates to a geodesic of simultaneity as in figure~\ref{FiguraTres}a.}
\label{FiguraQuatro2} 
\end{figure}

\ppparallel{\vspace{2em}
\ppl{\section*{~}\vspace{-1em}} \nopagebreak
\ppr{\section*{References}\vspace{-1em}} \ppp \nopagebreak
\vspace{-1.9em}}
\selectlanguage{esperanto}

\ppparallel{\end{Parallel}}

\begin{thebibliography}{20}
\selectlanguage{english}

\bibitem{reltemp2} F.M. Paiva, A.F.F. Teixeira, {\it The relativistic time -- II / La relativeca tempo -- II}, balda\u u aperonta en arXiv[physics]. 

\bibitem{Anderson} J.L. Anderson, {\it Principles of relativity physics}, Academic Press (1967).

\bibitem{Schwarzschild} F.M. Paiva, A.F.F. Teixeira, {\it Doppleraj efikoj \^ce Schwarzschild / Doppler effects in Schwarzschild}\,,  arXiv:0912.1229; CBPF-NF-023/09. 

\end{thebibliography}
\end{document}